\documentclass[final]{svjour3}
\usepackage[dvipdfmx]{graphicx}
\usepackage{rotating}
\usepackage{amssymb}
\usepackage{mathptmx}

\usepackage[numbers]{natbib}

\usepackage{xcolor}
\makeatletter
\journalname{Journal of Low Temperature Physics}
\usepackage{hyperref}

\begin{document}

\newcommand{\hdblarrow}{H\makebox[0.9ex][l]{$\downdownarrows$}-}
\title{The KISS experiment}

\author{A. Fasano$^1$ \and M. Aguiar$^2$ \and A. Benoit$^1$ \and A. Bideaud$^1$ \and O. Bourrion$^3$ \and M. Calvo$^1$ \and A. Catalano$^3$ \and A. P. de Taoro$^2$ \and G. Garde$^1$ \and A. Gomez$^4$ \and M. F. Gomez Renasco$^2$ \and J. Goupy$^1$ \and C. Hoarau$^3$ \and R. Hoyland$^2$ \and J. F. Mac\'ias-P\'erez$^3$ \and J. Marpaud$^3$ \and A. Monfardini$^1$ \and G. Pisano$^5$ \and N. Ponthieu$^6$ \and J. A. Rubi\~no Mart\'in$^2$ \and D. Tourres$^3$ \and C. Tucker$^5$ \and A. Beelen$^7$ \and G. Bres$^1$ \and M. De Petris$^8$ \and P. de Bernardis$^8$ \and G. Lagache$^7$ \and L. Lamagna$^8$ \and G. Luzzi$^8$ \and M. Marton$^3$ \and S. Masi$^8$ \and R. Rebolo$^2$ \and S. Roudier$^3$}

\institute{\email{alessandro.fasano@neel.cnrs.fr}}
\institute{
	$^1$ Institut N\'eel, CNRS and Universit\'e Grenoble Alpes, 25 av. des Martyrs, F-38042
	Grenoble, France\\ 
	$^2$ Instituto de Astrof\'sica de Canarias, V\'ia L\'actea s/n, E-38205 La Laguna, Tenerife, Spain \\ 
	$^3$ Univ. Grenoble Alpes, CNRS, Grenoble INP, LPSC-IN2P3, 53, avenue des Martyrs, 38000 Grenoble, France \\ 
	$^4$ Centro de Astrobiologia (CSIC-INTA), Madrid, Spain \\ 
	$^5$ Astronomy Instrumentation Group, University of Cardiff, UK \\ 
	$^6$ Univ. Grenoble Alpes, CNRS, IPAG, F-38000 Grenoble, France   \\ 
	$^7$ Aix Marseille Universit\'e, CNRS, LAM (Laboratoire d'Astrophysique de Marseille) UMR 7326, 13388, Marseille, France \\ 
	$^8$ Dipartimento di Fisica, Sapienza Universit\`a di Roma, Piazzale Aldo Moro 5, I-00185 Roma, Italy}

\maketitle

\begin{abstract}

Mapping millimetre continuum emission has become a key issue in modern multi-wavelength astrophysics. In particular, spectrum-imaging at low frequency resolution is an asset for characterizing the clusters of galaxies via the Sunyaev Zel'dovich (SZ) effect. In this context, we have built a ground-based spectrum-imager named KIDs Interferometer Spectrum Survey (KISS). This instrument is based on two 316-pixel arrays of Kinetic Inductance Detectors (KID) cooled to 150 mK by a custom dilution refrigerator-based cryostat. By using Ti-Al and Al absorbers, we can cover a wide frequency range between 80 and 300 GHz. In order to preserve a large instantaneous Field of View (FoV) $\sim1^\circ$ the spectrometer is based on a Fourier Transform interferometer. This represents a technological challenge due to the fast scanning speed that is needed to overcome the effects of background atmospheric fluctuations. KISS is installed at the QUIJOTE 2.25 m telescope in Tenerife since February 2019 and is currently in its commissioning phase. In this proceeding we present an overview of the instrument and the latest results.


\end{abstract}

\section{Scientific Motivation}

The	determination of the distribution of cluster of	galaxies as	a function of mass and redshift has	been proven	to be a	powerful cosmological tool, in particular to infer the matter content of the universe and its dynamics. These objects have been traditionally observed at optical and X-ray wavelengths, in this last case via bremsstrahlung emission of the hot (from 10$^7$ up to 10$^8$ K) plasma that composes the Intra Cluster Medium (ICM). However, thanks to the fast evolution of millimetre and sub-millimetre instruments it is now common to  observe clusters of galaxies via the SZ effect. The SZ effect is a spectral distortion of the Cosmic Microwave Background (CMB) and is due to the Inverse-Compton scattering of the photons interacting with the ICM. Two main effects can be considered: the thermal SZ (tSZ), which refers to the interaction of the hot electrons in clusters with the CMB	photons, and the kinetic SZ (kSZ), which is a Doppler shift induced by the relative motions within the cluster (see \cite{sz} for the detailed description).

The complex morphology can introduce significant bias in the relationship between tSZ flux and mass.	Astrophysical contamination by local and foreground	emission (e.g. thermal dust, synchrotron, dusty and radio galaxies) limits	the	accuracy of	the	reconstruction	of	the	tSZ	effect (as described in detail in \cite{ruppin}).	The	former	can	be addressed with continuum dual-band observations at high angular resolution	such as	those performed e.g. with NIKA \cite{nika} and NIKA2 \cite{adam2018}. For the	latter,	multi-wavelength observations are needed to	separate the different components and to add complementary information.

In this context, we have built KISS, a ground-based spectrum-imager dedicated to the study of clusters of galaxies at millimetre wavelengths, where the SZ effect is the most prominent. Compared to the predecessors that are imagers (e.g. ACT \cite{Kosowsky}, Mustang \cite{mustang}, SPT \cite{Staniszewski}, Bolocam \cite{bolocam}, the satellite Planck \cite{planck2016}, NIKA \cite{nika} and NIKA2 \cite{adam2018}), KISS is a Fourier Transform Spectrometer (FTS) and allows for accurate component separation for low redshift clusters.

\section{KISS instrument}

\subsection{The detectors: KIDs}

\begin{figure}[htbp]
	\begin{center}
		\includegraphics[width=0.8\linewidth, keepaspectratio]{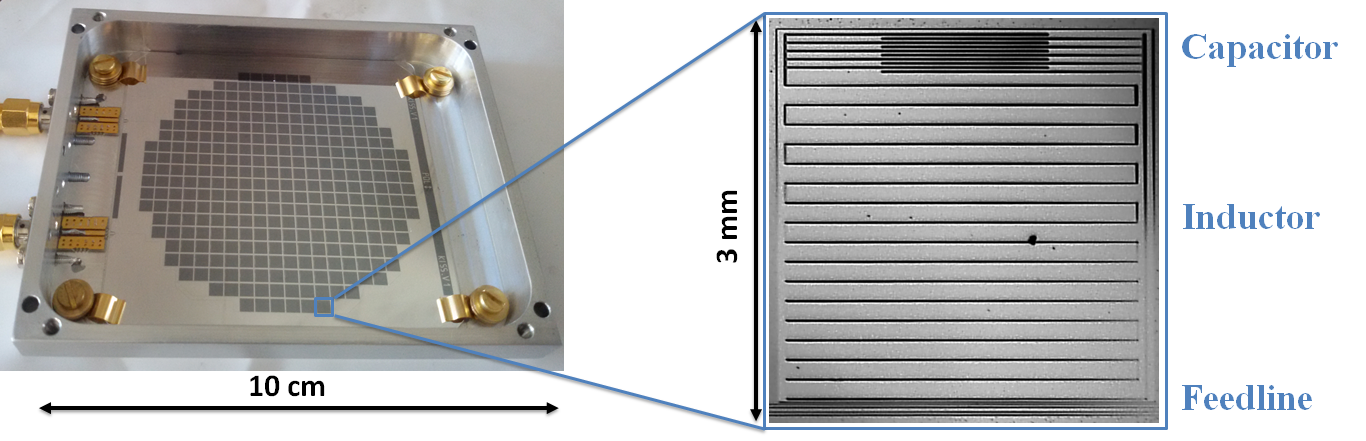}
		\caption{Left: array of KISS with a total of 316 pixel. Right: zoom over a single pixel, we can notice the polarised meander geometry.}
		\label{fig:array}
	\end{center}

\end{figure}

The exploitation of a camera capable to cover a wide frequency range, as mentioned before, is due to the necessity to discriminate the different components of the SZ effect as well as the precise subtraction of the contaminations. The technology selected for this purpose is the KID (see \cite{kids} for a review): a technology already used in experiments such as  NIKA, NIKA2, MUSIC \cite{music}, A-MKID \cite{amkid} and OLIMPO \cite{paiella} and planned for a number of others (e.g. MUSCAT and Toltec). In Fig.~\ref{fig:array} we see one of the two KISS arrays, with a zoom over a single pixel.

KIDs are today a natural option for millimetre wavelengths instrumentation, as first demonstrated by NIKA. Working at a very low and stable temperature allows one to neglect all the sources of noises that strictly depend on the temperature and to achieve photon-noise limited conditions, i.e. the detectors noise depends only on the flux of background photons. A KID consists in a high quality factor resonator working in the superconducting regime at $T\ll T_c$, i.e., well below the critical temperature ($T_c$). In KISS, the 150 mK working temperature is reached through the combination of active cooling elements composed of a Pulse-Tube cryocooler and a custom $^3$He/$^4$He dilution refrigerator. Within this family of detectors, we adopted the lumped-elements configuration. In this configuration the design of the resonator ensures that the current distribution in the meander is homogeneous leading to an isotropic absorption of the incident photons. The frequency of resonance is a physical quantity: it can be recorded and converted to an input optical power. In a superconductor, the electron populations split into free electrons and Cooper pairs, which are pairs of electrons bound together by an attractive force originating between the lattice and the quasi-particles. The KID is sensitive to changes in the Cooper-pair population which is directly related to the resonance frequency through the kinetic inductance.

KIDs have an intrinsic low frequency cut-off directly related to the $T_c$ of the absorber but, theoretically, they can reach arbitrarily high frequency. In the case of KISS, we adopted 22 nm thick Al absorber, with a minimum detectable frequency of 110 GHz. For the future,	 we will implement the Ti-Al bylayer technology \cite{bilayer} which will enable observations down to 80 GHz. The upper frequency is constrained by the optical filtering to 300 GHz. KISS has two polarisation-sensitive 316-pixel arrays, each covering the whole spectral band and simultaneously mapping the same portion of the sky.

\subsection{Spectroscopy with the Martin-Puplett Interferometer}

As mentioned in the introduction, key issues for future cosmological experiments are:

\begin{itemize}
	\item multi-wavelength observations;
	\item high mapping speed $\left( v_{speed}\propto\frac{FoV}{NEP^2}\right)$, where $NEP$ is the Noise Equivalent Power. 
\end{itemize}

\noindent Thus, the challenge is to develop multi-frequency instruments with large FoV. In this context, a natural candidate is represented by FTS and, in particular, by the Martin-Puplett Interferometer (MPI), adopted for KISS. The drawback for this class of spectrometers is the larger photon noise: the single pixel receives the whole electromagnetic band. The main alternatives are the on-chip dispersive spectroscopy (e.g. see \cite{deshima}) and the grating spectroscopy \cite{grating}. In which each detector is dedicated to a single frequency. This translates in a lower photon noise but it does not allow large FoV because of the prohibitive pixel counts. This technology is still under development and optimization thus far for the on-chip.

\begin{figure}[htbp]
	\begin{center}
		\includegraphics[width=0.9\linewidth, keepaspectratio]{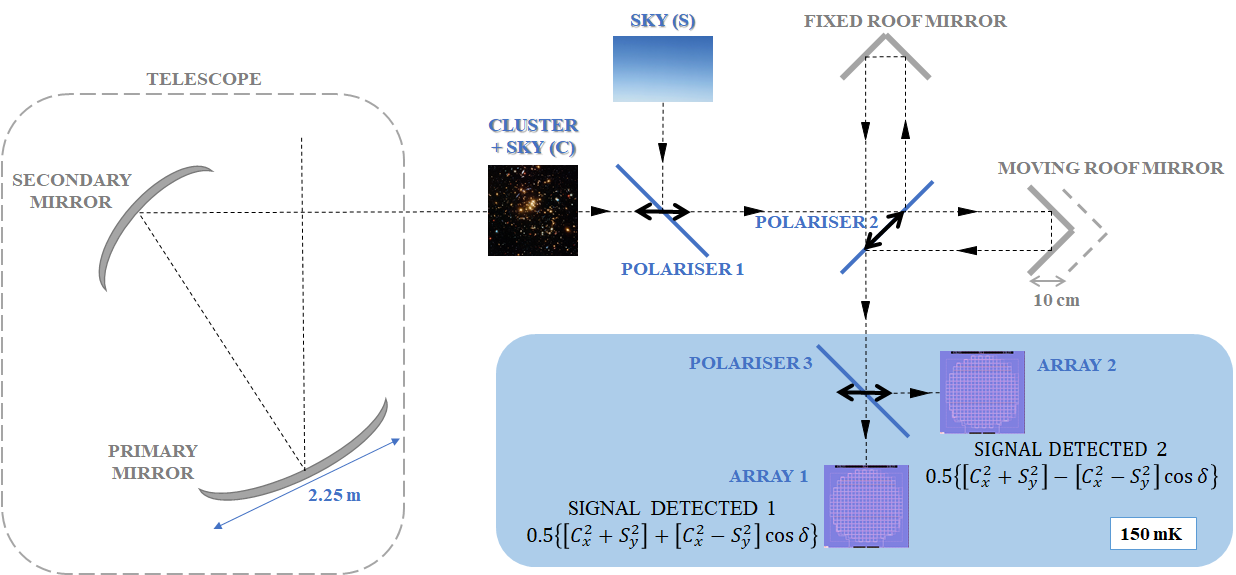}
		\caption{KISS experiment scheme. Left: QUIJOTE telescope. Right: the optical path goes through the MPI reaching the two arrays on the focal plane at 150 mK.}
		\label{fig:sketch}		
	\end{center}
\end{figure}

A schematic view of the KISS instrument and its MPI is represented in Fig.~\ref{fig:sketch}. The MPI measures the difference in power between two input beams, as described in \cite{mpi} \textcolor{black}{and, as already demonstrated by the FIRAS instrument \cite{firas}, it is a reliable technique to perform precise measurements of CMB spectrum. In the KISS instrument the two input beams are represented in Fig.~\ref{fig:sketch} by the FoV (cluster+sky) and a 3-degree defocused portion of the sky that includes the FoV (sky). In addition, a tilting mirror permits to switch to another configuration where the secondary source (sky) is represented by a mirror that looks to a stable-temperature cryostat stage. In both secondary-source configurations the} two signals are split and are later recombined through polarisers. The multi-frequency behaviour is granted by a moving roof mirror that introduces an Optical Path Difference (OPD) and creates the interference. The range of the mirror displacement defines the spectral resolution. The output signal is then analysed by the Fourier Transform technique that produces the spectrum.
Analytically, we know that the two outputs, expressed in power, are:

\begin{equation}
I^{1,2}_{\nu}(\delta) = \frac{1}{2}\left[ C^2_x+S^2_y \right]\pm \frac{1}{2}\left[ C^2_x-S^2_y \right]\cos(\delta)
\end{equation}

\noindent where $\delta\equiv \omega \cdot t=2\pi\nu\cdot \Delta x /c $, $\delta$ is the phase difference, $\nu$ the frequency, $\Delta x$ twice the displacement of the moving mirror, $c$ the light speed, $C_x$ and $S_y$ are the powers emitted by two input sources as shown in Fig.~\ref{fig:sketch} (selected in polarisation by the first polariser). The plus-minus sign identifies the two different outputs.

The fast sweep of the  moving mirror makes possible to study the signal under stable background conditions. This is fundamental for the ground-based observations as it is needed to overcome the effects of the atmospheric fluctuations, \textcolor{black}{We expect such fluctuations to have a 1/f knee around $\sim$1 Hz for KISS. The subtraction of atmosphere contribution is guaranteed by the defocused sky configuration as a secondary source and, in addition, we can further study it comparing the results with the two different secondary sources.}

\begin{figure}[htbp]
	\begin{center}
		\includegraphics[width=0.7\linewidth, keepaspectratio]{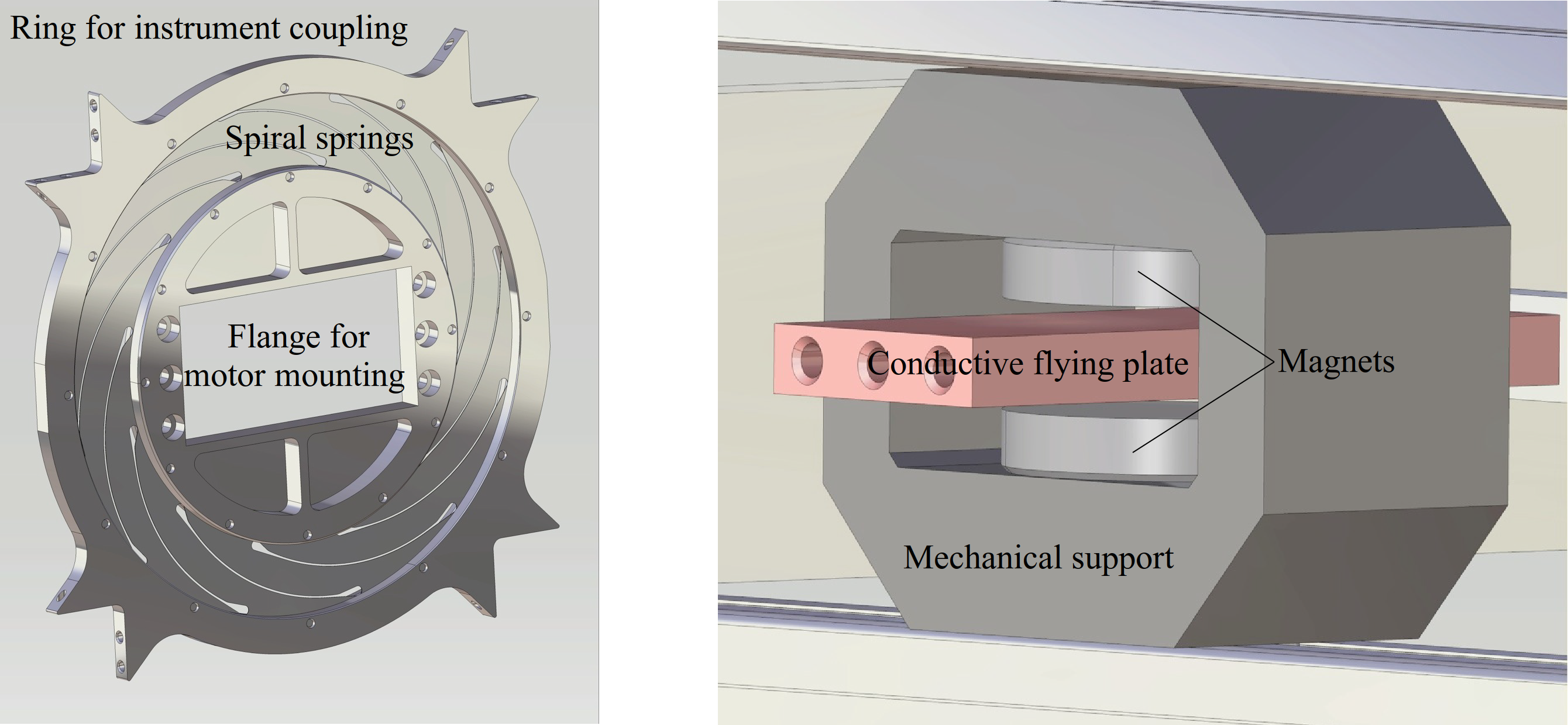}
		\caption{Left: anti-vibration system based on spiral springs. Right: eddy current brake designed to dump residual velocity. \textcolor{black}{ The moving copper plate experiences a drag force from the magnet that opposes its motion.}}
		\label{fig:antivibrations}		
	\end{center}
\end{figure}

The MPI technique introduces a major issue on the instrument development: the fast engine oscillations (5 Hz) need to be stable. The first element introduced is a counter-engine: having a motion opposite to that of the mirror engine, it nullifies the total inertial momentum. Secondly, we adopted spiral springs to decouple the residual vibrations from the focal plane (Fig.~\ref{fig:antivibrations}). This solution is derived from the Planck Joule-Thompson cryogenic machine (see \cite{thermal} for more information). Finally, the slower accelerations are damped by an eddy current brake placed between the two engines. Thanks to this setup, we are not polluted with vibrations on the focal plane.

\section{Sensitivity and first light}

\begin{figure}[htbp]
	\begin{center}
		\includegraphics[width=0.6\linewidth, keepaspectratio]{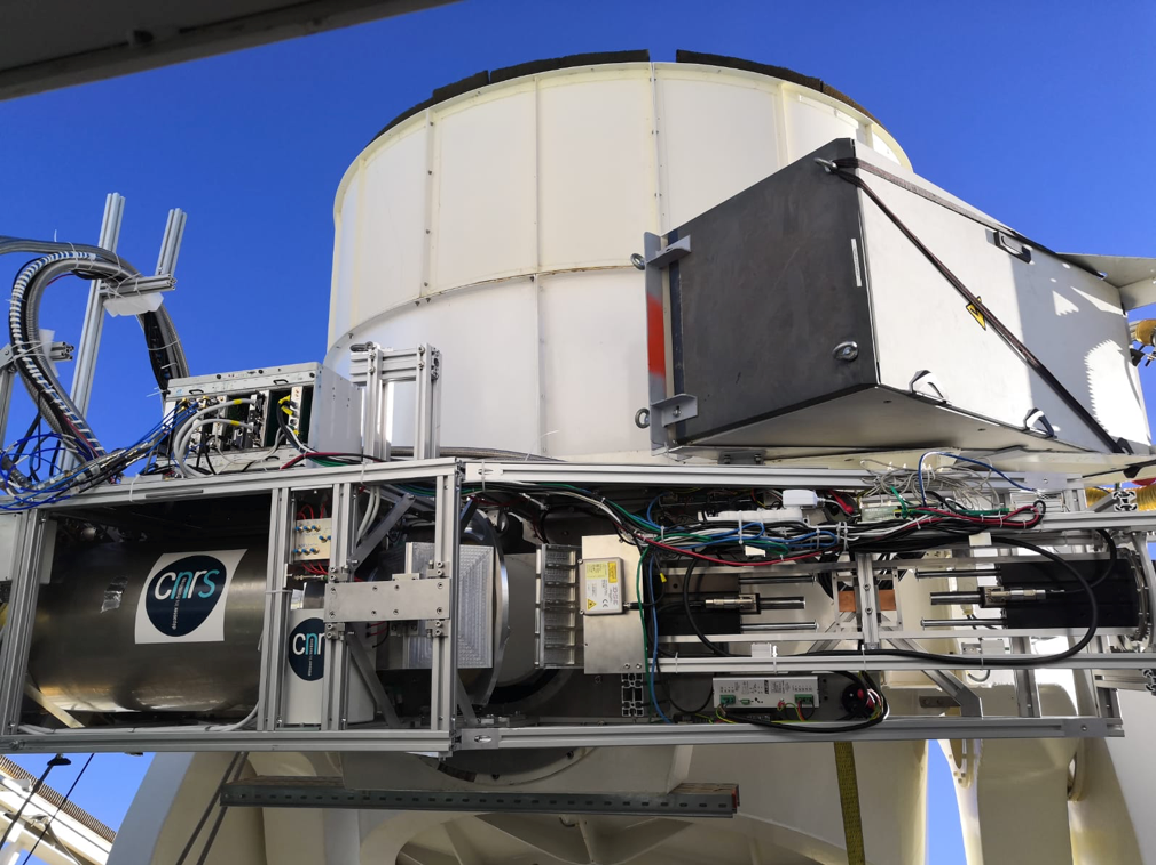}
		\caption{KISS instrument installed at the QUIJOTE telescope in Tenerife.}
		\label{fig:kiss}	
	\end{center}
\end{figure}

We describe in this section the preliminary results of our commissioning campaigns. KISS has been installed at the QUIJOTE telescope in February 2019. Fig.~\ref{fig:kiss} shows it integrated at the telescope.

\begin{figure}[htbp]
	\begin{center}
		\includegraphics[width=0.65\linewidth, keepaspectratio]{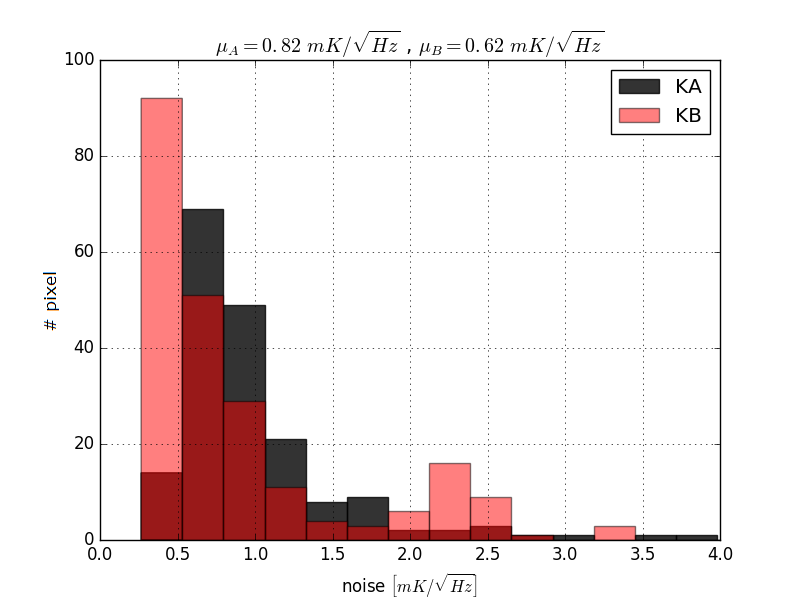}
		\caption{Single pixel Noise Equivalent Temperature histogram. The units in mK are Rayleigh-Jeans corrected for the source temperature and are for the photometric mode, i.e. for the whole frequency range. The two \textcolor{black}{shaded} colours identify the arrays, named ``KA'' and ``KB''. The single pixel has a beam of 4 arcmin at 150 GHz. \textcolor{black}{75\% of the pixels are present because the optimisation of the single-array read out at 312 is still under optimisation.}}
		\label{fig:NET}		
	\end{center}
\end{figure}

After coupling the instrument to the telescope, the first step was the qualification on the sky, i.e., comparing its performances to the ones established in the laboratory. In Fig.~\ref{fig:NET} we show the Noise Equivalent Temperature (NET) measured from on-sky observations, which show perfect agreement with the laboratory tests previously performed. The minimum detectable flux at fixed signal-to-noise ratio and integration time directly depends on the NET, which is proportional to the NEP. 

\begin{figure}[htbp]
	\begin{center}
		\includegraphics[width=1.\linewidth, keepaspectratio]{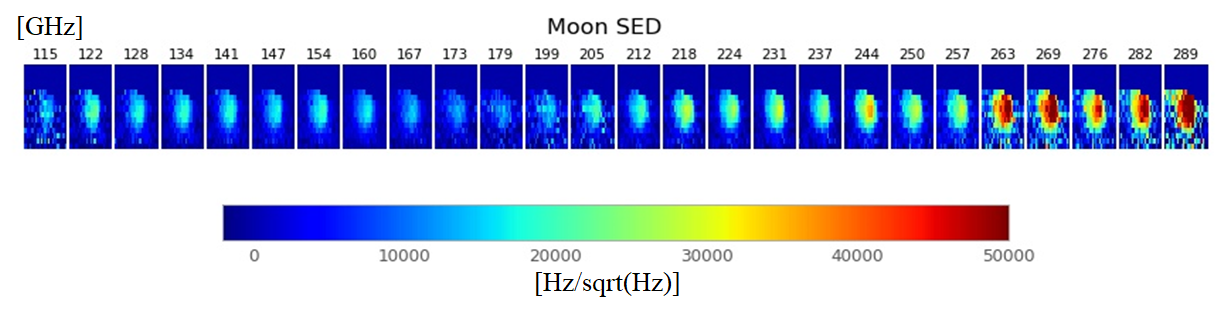}
		\caption{Un-calibrated multi-wavelength maps with 6 GHz bin and transmission band convoluted.}
		\label{fig:maps}			
	\end{center}
\end{figure}

The following step was the demonstration of the multi-wavelength mapping capability, which can be considered the KISS ``first light''. As shown in Fig.~\ref{fig:maps}, KISS can produce maps at different wavelengths simultaneously. In this particular case, we focused on Moon observation.

\noindent The Moon radiance required the installation of a diaphragm to reduce the input signal, the presence of which made it difficult to perform an absolute calibration. The maps shown here are, therefore, not calibrated and the intensity reported is the signal amplitude spectral density in $Hz/\sqrt{Hz}$, where the $Hz$ at the numerator represents the shift in the resonance frequency. The shift in frequency can be converted to incident optical power (\cite{swenson}, \cite{monfa}). In the maps, we can see the transmission cut-off the notch filter at 180 GHz, which as been adopted to avoid contamination from the H$_2$O atmospheric absorption line.

The point source calibration (e.g. Jupiter) and the beam characterisation are in progress; we will present the results in a future publication. With these we will be able to fully characterise the instrument and calibrate the maps. If successful, these steps will be followed by the first cluster of galaxies long integration. Besides the intrinsic scientific interest, this will also pave the way for the future multi-wavelength observation at millimetre wavelengths. 

We plan to exploit KID array based on Ti-Al bilayer. This will make the lower part of the 80-300 GHz band accessible to KISS, as well as representing the first on-sky performance test of a technology validated only in the laboratory up until now.

\bibliography{bibliography.bib}

\end{document}